\documentclass[aps,prl,twocolumn,showpacs,10pt,superscriptaddress,preprintnumbers,nofootinbib]{revtex4-1}
\usepackage{epsfig,amssymb,amsmath,psfrag,epstopdf,color}
\pdfoutput=1
\usepackage{graphicx}
\usepackage{subcaption}
\usepackage{ragged2e}
\DeclareCaptionJustification{justified}{\justifying}
\usepackage[normalem]{ulem}
\usepackage{paralist}
\usepackage{hyperref}
\usepackage[size=tiny]{todonotes}
\allowdisplaybreaks

\def\beq{\begin{equation}}
\def\eeq{\end{equation}}
\def\bsp#1\esp{\begin{split}#1\end{split}}
\newcommand{\be}{\begin{equation}}
\newcommand{\ee}{\end{equation}}
\newcommand{\bea}{\begin{eqnarray}}
\newcommand{\eea}{\end{eqnarray}}

\def\Fig#1{Fig.~{\ref{#1}}}

\def\to{\rightarrow}

\newcommand{\comment}[1]{}

\newcommand{\td}{\mathrm{d}}
\newcommand{\SO}[1]{\mathrm{SO}(#1)}

\newcommand{\TT}[1]{\mathrm{#1}}

\renewcommand{\>}{\right\rangle}
\newcommand{\<}{\left\langle}

\newcommand{\As}{\alpha_{\mathrm{s}}}

\newcommand{\cE}{\mathcal{E}}
\newcommand{\LQCD}{\Lambda_{\mathrm{QCD}}}

\begin{document}

\title{Resolving the Scales of the Quark-Gluon Plasma with Energy Correlators}


\author{Carlota Andres}
\affiliation{CPHT, CNRS, Ecole polytechnique, IP Paris, F-91128 Palaiseau, France}

\author{Fabio Dominguez}
\affiliation{Instituto Galego de F{\'{i}}sica de Altas Enerx{\'{i}}as (IGFAE),  Universidade de Santiago de Compostela, Santiago de Compostela 15782,  Spain}

\author{Raghav Kunnawalkam Elayavalli}
\affiliation{Wright Laboratory, Yale University, New Haven, CT}
\affiliation{Brookhaven National Laboratory, Upton NY}
\affiliation{Department of Physics and Astronomy, Vanderbilt University, Nashville, TN}

\author{Jack Holguin}
\affiliation{CPHT, CNRS, Ecole polytechnique, IP Paris, F-91128 Palaiseau, France}

\author{Cyrille Marquet}
\affiliation{CPHT, CNRS, Ecole polytechnique, IP Paris, F-91128 Palaiseau, France}

\author{Ian Moult}
\affiliation{Department of Physics, Yale University, New Haven, CT 06511}

\begin{abstract}
Jets provide us with ideal probes of the quark-gluon plasma (QGP) produced in heavy-ion collisions, since its dynamics at its different scales is imprinted into the  multi-scale substructure of the final state jets.
We present a new approach to jet substructure in heavy-ion collisions based on the study of correlation functions of energy flow operators. 
By analysing the two-point correlator of an in-medium quark jet, we demonstrate that the spectra of correlation functions robustly identify the scales defined by the properties of the QGP, particularly those associated with the onset of colour coherence. 
\end{abstract}

\maketitle


\emph{Introduction.}---The quark-gluon plasma (QGP) is an extreme state of quantum matter, whose study provides insights into the phase structure of quantum chromodynamics (QCD), the dynamics of free quarks and gluons, the mechanism of hadronisation, and, more generally, the dynamics and properties of relativistic strongly interacting matter. The ability to create the QGP in ultra-relativistic heavy-ion collisions at the Relativistic Heavy Ion
Collider (RHIC) and at the Large Hadron Collider (LHC) \cite{PHOBOS:2004zne,Muller:2006ee,Muller:2012zq} has led to impressive theoretical and experimental developments aiming at its description. For recent reviews see~\cite{Connors:2017ptx,Busza:2018rrf,Cunqueiro:2021wls,Apolinario:2022vzg}.

\begin{figure}
\includegraphics[width=0.40\textwidth]{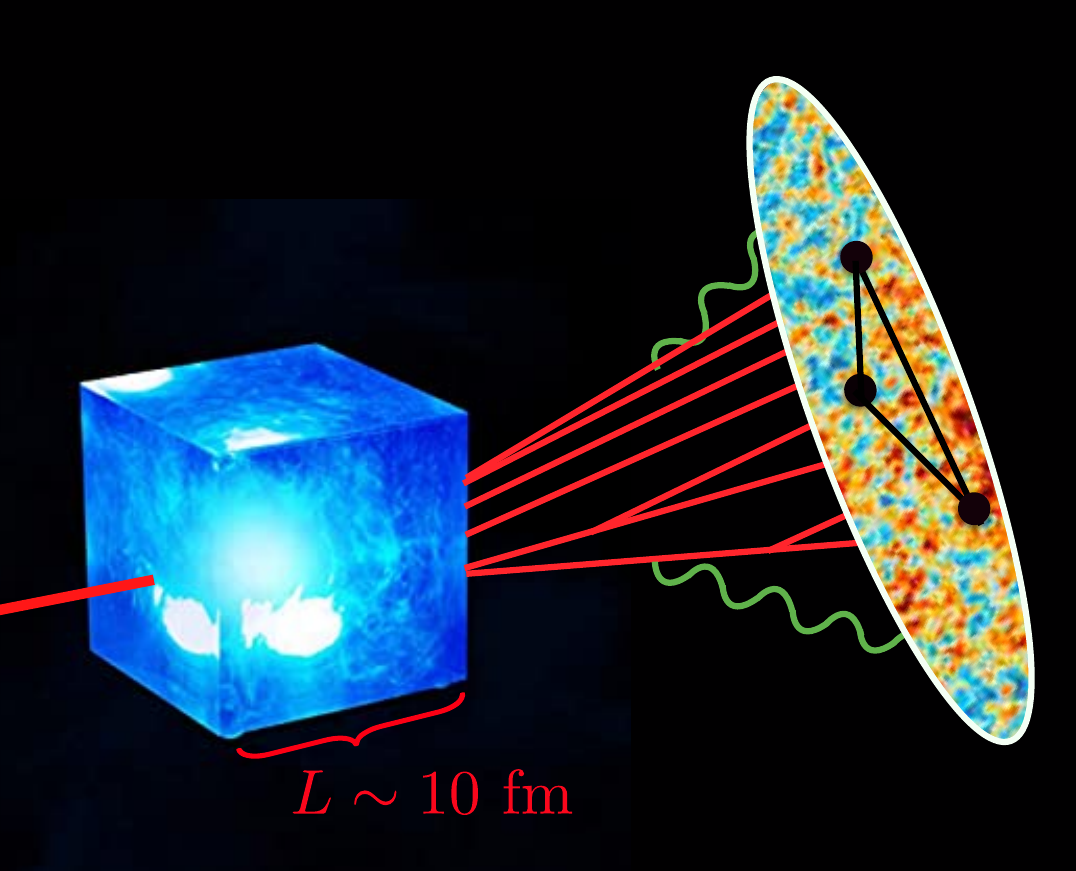}
  \caption{The scales of the QGP are imprinted into the scales of the substructure of final state jets. These can be naturally extracted through the use of correlation functions of energy flux.}
  \label{fig:picture}
\end{figure}

In analogy to how the classic Geiger-Marsden experiments discovered the structure of the atom \cite{doi:10.1080/14786440408634197}, high-energy quarks and gluons produced in the hard scattering process travel through the QGP, and provide natural probes of its structure. However, due to the parton shower and confinement process, the final state observed in a heavy-ion experiment is not a single parton, but rather a jet, namely a complicated multi-scale object, as illustrated in \Fig{fig:picture}. While the jet nature of the final state complicates its interpretation, it also presents a remarkable opportunity, since the dynamics at different scales in the QGP are imprinted in the different scales of the substructure of the jet. For this reason, jet substructure, an active area of research in p$+$p collisions at the LHC \cite{Dasgupta:2013ihk,Larkoski:2013eya,Larkoski:2017jix,Asquith:2018igt,Marzani:2019hun}, has attracted significant interest from the heavy-ion community, see e.g. \cite{Andrews:2018jcm,Cunqueiro:2021wls,CMS:2013lhm,CMS:2018zze,Chien:2015hda,Chien:2016led,Apolinario:2017qay,Ringer:2019rfk,Vaidya:2020lih,Caucal:2021cfb,Mehtar-Tani:2021fud,Milhano:2022kzx}. However, the subtle manner in which the QGP properties are imprinted into standard jet substructure observables has made it difficult to extract robust conclusions about the dynamics of the QGP. These subtleties are due in part to the complicated experimental environment where the large fluctuating background from the underlying event inhibits clean measurements; and in part due 
to the presence of multiple phenomena such as colour coherence \cite{Mehtar-Tani:2010ebp,Mehtar-Tani:2011hma,Casalderrey-Solana:2011ule,Mehtar-Tani:2012mfa} and medium response \cite{Cao:2020wlm,CMS:2021otx}, which have hampered a clear theoretical interpretation of traditional jet observables.

In this \emph{Letter} we present a novel approach to jet substructure in heavy-ion collisions formulated in terms of correlation functions. Our method is based on the insight that the features of the QGP are clearly imprinted at specific time scales in the jet, as opposed to through the modification of jets of a fixed shape, as previously considered in the literature. Our goal is to isolate these scales in the correlator spectra and thus disentangle the different properties of the QGP. Since collider experiments only make measurements at asymptotic infinity, the most appropriate collider correlators are correlation functions of flux operators $\langle\mathcal{E}(\vec{n}_{1}) \dots \mathcal{E}(\vec{n}_{k})\rangle $ \cite{Basham:1979gh,Basham:1978zq,Basham:1978bw,Basham:1977iq,Hofman:2008ar} where $\mathcal{E}(\vec{n}_{1})$ measures the asymptotic energy flux  in the direction $\vec{n}_{1}$ \cite{Ore:1979ry,Korchemsky:1999kt,Hofman:2008ar,Belitsky:2013xxa}. Due to recent theoretical progress in the understanding of these operators \cite{Kravchuk:2018htv}, their operator product expansion (OPE) \cite{Hofman:2008ar,Kravchuk:2018htv,Kologlu:2019bco,Kologlu:2019mfz,Chang:2020qpj,Chen:2021gdk,Chen:2022jhb,Chang:2022ryc}, and their perturbative structure \cite{Belitsky:2013ofa,Dixon:2018qgp,Chen:2019bpb,Henn:2019gkr,Yan:2022cye,Yang:2022tgm}, it is now possible to directly use energy correlators for the phenomenology of jet substructure \cite{Komiske:2022enw,Chen:2022swd,Holguin:2022epo,Lee:2022ige,Ricci:2022htc,Liu:2022wop,Holguin:2022epo}.

We propose here the first application of multi-point correlation functions to jets in heavy-ion collisions, allowing sensitivity to the internal \emph{angular} scales within the jets. We are motivated to do so because, despite best efforts, present approaches to jet substructure based on grooming and algorithmic de-clustering \cite{Larkoski:2015lea,Larkoski:2017bvj} have been unable to introduce a robust \emph{angular} variable for jets in heavy-ion collisions, due in part to grooming techniques not having been tailored to the presence of a large underlying event which leads to a significant number of splittings to be misidentified \cite{Mulligan:2020tim}. In contrast, the angular scales accessed by energy correlators have a fundamentally different response to the realities of the experimental environment. In vacuum, energy correlators display a power-law behaviour which is insensitive to both higher order corrections and non-perturbative effects \cite{Chen:2022jhb,Komiske:2022enw}. Their study in heavy-ion processes is especially enticing since, due to the energy weighting, they are broadly insensitive to soft physics, and thus background and medium-response effects are expected to be subleading even without the application of grooming techniques. Additionally, energy correlators can be computed on tracks \cite{Chen:2020vvp,Li:2021zcf,Jaarsma:2022kdd,Chang:2013rca,Chang:2013iba}, which allows for higher angular resolution and suppresses pile-up. 

As a proof of concept, we analyse here the specific case of the two-point correlator $\langle \cE(\vec n_1) \cE(\vec n_2) \rangle$, which introduces a single scale sensitive angular parameter, $\cos \theta =\vec n_1 \cdot \vec n_2 $. We compute the two-point correlator (EEC) on a quark jet which propagates through a static medium of finite length within the BDMPS-Z formalism~\cite{Baier:1996kr,Baier:1996sk,Zakharov:1996fv,Zakharov:1997uu}. We find that the correlator allows us to identify the onset of colour coherence on the quark-gluon splitting. This can be expressed in terms of the emergence of a \emph{resolution scale} for the QGP, $\theta_c$, an intrinsic medium angular scale that divides the radiation phase-space into resolved and unresolved splittings~\cite{Casalderrey-Solana:2012evi,Mehtar-Tani:2017ypq}. Furthermore, we show that the key features we identify, embodied in enhancement at wide-angles, are also present in the EEC spectrum computed on in-medium jet events produced by the JEWEL event generator~\cite{Zapp:2008gi,Zapp:2012ak,Zapp:2013vla}.


\emph{The Two-Point Correlator.}--- The $n$-th  weighted normalised two-point correlator can be computed from the inclusive cross-section ($\sigma_{ij}$)  to produce two hadrons ($i,j$) as %
\begin{align}
    &\frac{\< \cE^{n}(\vec n_1) \cE^{n}(\vec n_2) \>}{Q^{2n}} \\  
    &=\frac{1}{\sigma} \sum_{ij} \int \frac{\td \sigma_{ij}}{\td \vec n_i \td \vec n_j } \frac{E^{n}_{i} E^{n}_{j}}{Q^{2n}} \delta^{(2)}(\vec n_i - \vec n_{1})\delta^{(2)}(\vec n_j - \vec n_{2})\,, \nonumber
\end{align}
where $E_{i}$ is the lab-frame energy of hadron $i$, and $Q$ is an appropriate hard scale. In an isotropic environment, 3 of the degrees of freedom in $\{\vec n_1,\vec n_2\}$ correspond $\SO{3}$ symmetries. Thus we will study the distribution
\begin{align}
    &\frac{\td \Sigma^{(n)}}{\td \theta} = \int \td \vec n_{1,2} \frac{\< \cE^{n}(\vec n_1) \cE^{n}(\vec n_2) \>}{Q^{2n}} \delta(\vec n_2 \cdot \vec n_{1} - \cos \theta)\,. 
\end{align}
Strictly speaking, $\Sigma^{(n)}$ is only collinear-safe when $n=1$. However, for $n\in(1,2]$ the regular pattern of IR-divergences can be absorbed into moments of fragmentation functions or track functions~\cite{Chen:2020vvp,Li:2021zcf}. We will not consider $n>2$ in this \emph{Letter}.

In  vacuum, massless QCD is asymptotically conformal (up to the running coupling) and so obeys a celestial OPE~\cite{Hofman:2008ar} which can be used to show that $\Sigma^{(1)}$ displays a featureless power-law  behaviour \cite{Konishi:1979cb,Hofman:2008ar,Dixon:2019uzg,Kologlu:2019mfz}
\begin{align}
    \frac{\td \Sigma^{(1)}}{\td \theta} \sim \frac{1}{\theta^{1-\gamma(3)}} + \mathcal{O}(\theta^{0})\,,
\end{align}
where $\gamma(3)$ is the twist-2 spin-3 QCD anomalous dimension at fixed coupling\footnote{The running coupling slightly breaks the simple exponentiation of the anomalous dimension~\cite{Chen:2021gdk}, see \cite{Chen:2022jhb}.}. In hadron colliders, this power-law behaviour holds inside identified jets, as can be proven using a factorisation theorem \cite{Holguin:2022epo,Lee:2022ige} based on fragmenting jet functions \cite{Kang:2016mcy,Kang:2016ehg}, and has recently been observed in experimental LHC p$+$p data \cite{Komiske:2022enw}. For $\td \Sigma^{(2)}$ the result is more involved, but the leading-order scaling remains $\td \Sigma^{(2)}/\td \theta \sim \theta^{-1} + \mathcal{O}(\theta^{0},\As^2)$.

In contrast, radiation within a QCD jet formed whilst propagating through a medium becomes sensitive to scales defined by the medium properties. It has been argued that the medium cannot resolve emissions at arbitrarily small angles \cite{Casalderrey-Solana:2012evi,Mehtar-Tani:2017ypq,Dominguez:2019ges,Andrews:2018jcm}, thus introducing a resolution scale which takes the form of a minimal angle for medium-induced radiation. Consequently, we expect $\Sigma^{(n)}$, measured on a sample of medium jets, to mostly obey the perturbative vacuum behaviour below this minimum angle and display an excess above it\footnote{This medium scale is expected to be much larger than $\LQCD/Q$ where $\Sigma^{(n)}$ becomes sensitive to  hadronisation and displays the scaling properties of a free ideal gas of hadrons \cite{Komiske:2022enw}.}.

\begin{figure*}
\includegraphics[width=0.45\textwidth]{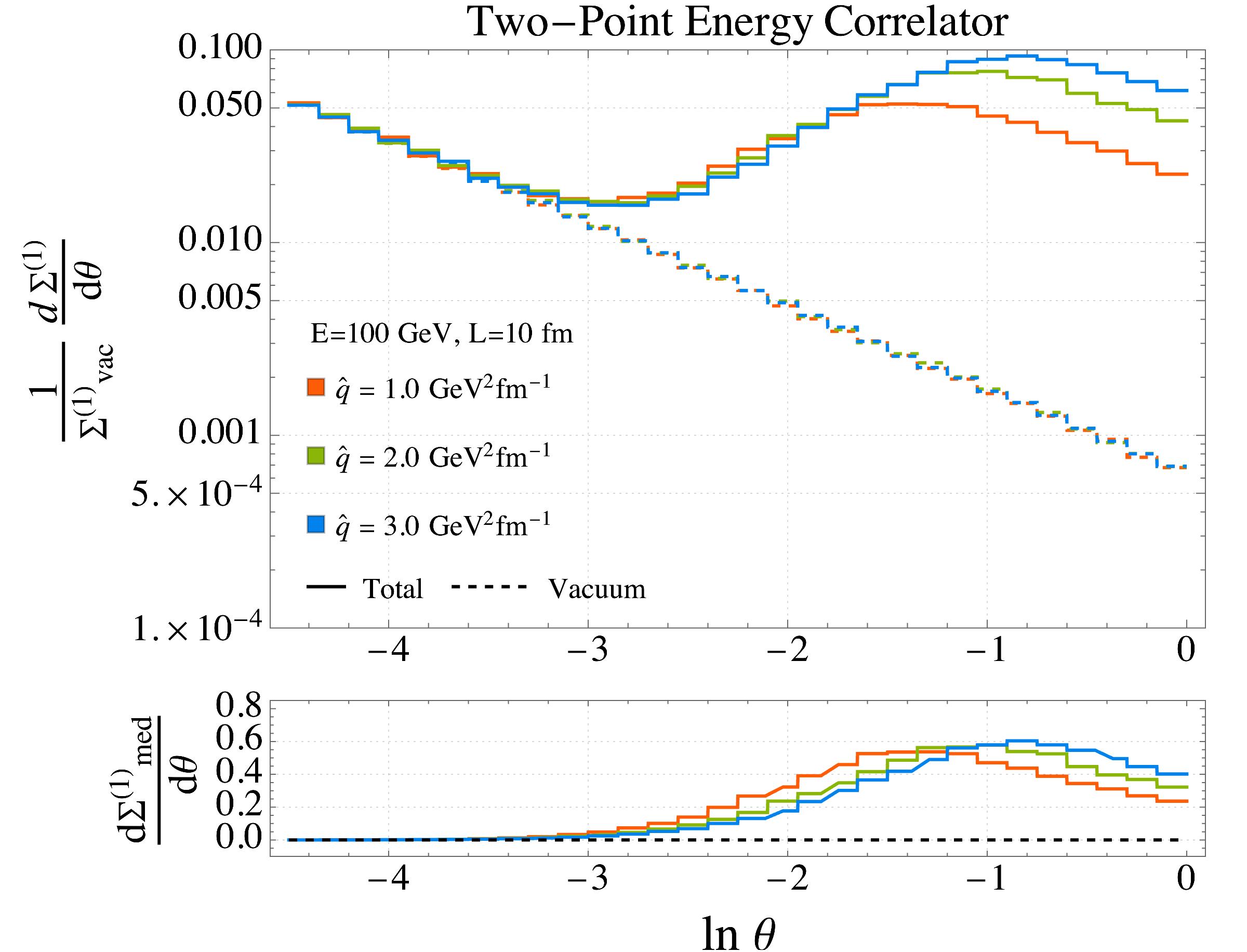}~~~~~~~\includegraphics[width=0.45\textwidth]{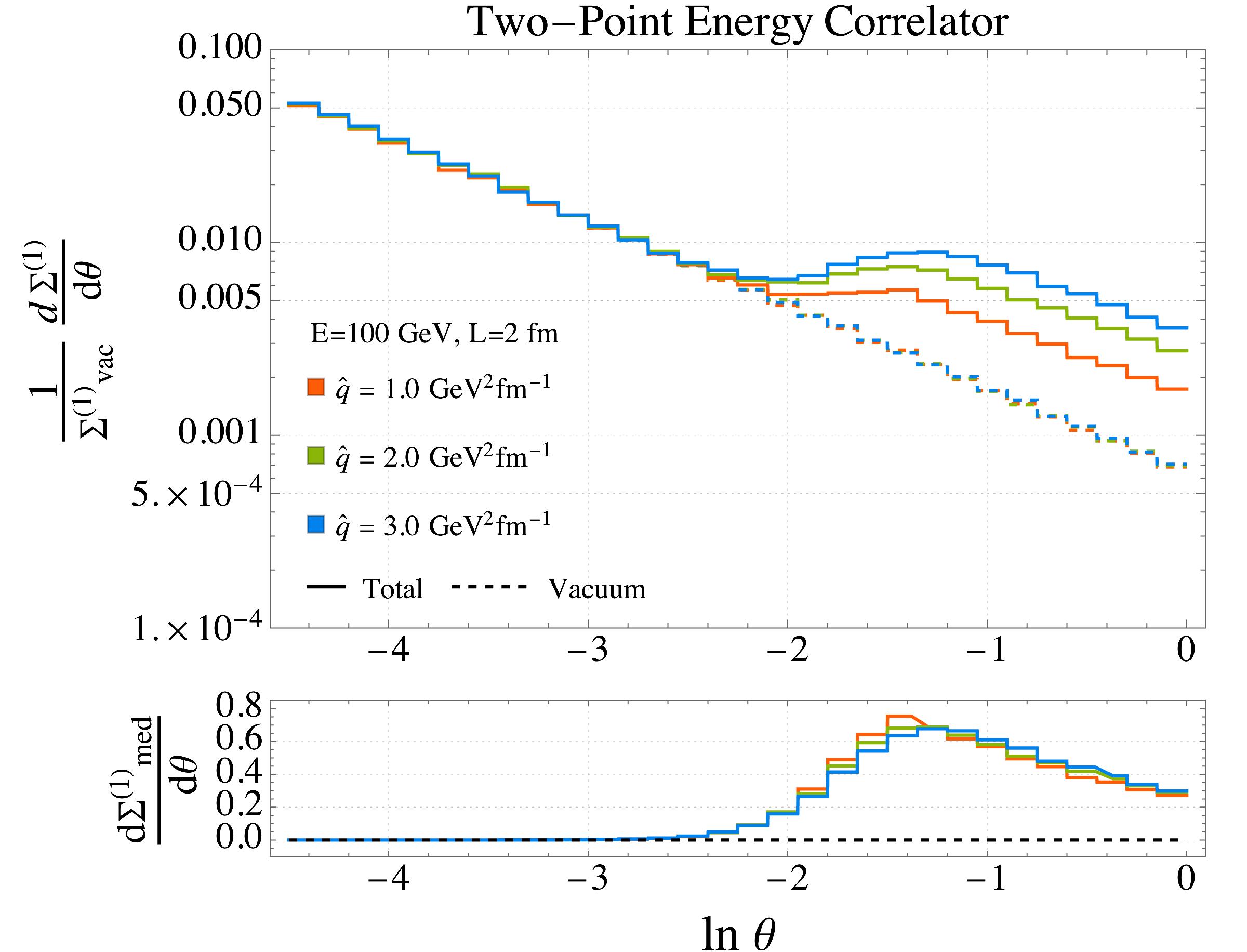}
\caption{The EEC evaluated using \eqref{eq:masterequation} for the DC (left panel) and PC regimes (right panel). The bottom panels show the volume normalised medium contribution to the distribution, defined as $\td \Sigma^{(n)}_{\TT{med}} = (\td \Sigma^{(n)} - \td \Sigma^{(n)}(\hat{q}=0))/\Sigma^{(n)}_{\TT{med}}$, so the shape can be more easily compared. The scales of the medium are clearly imprinted into the correlator.}
  \label{fig:EEC100GeV}
\end{figure*}

\emph{Analytical Framework.}---To illustrate the key features of the EEC, we consider a situation where a high-energy quark jet propagates through a medium. We are interested in the relatively wide angle region of medium modification where we can initially disregard the resummation of vacuum collinear sub-structure. Thus, to leading order in the semi-hard splittings we can compute $\td \Sigma^{(n)}$ as
\begin{align}
    \frac{\td \Sigma^{(n)}}{\td \theta} = \frac{1}{\sigma_{qg}}\int  \td z  \frac{\td \sigma_{qg}}{\td \theta \td z } z^n (1-z)^n + \mathcal{O}\left(\frac{\mu_{\TT{s}}}{E}\right)\,,
\end{align}
where $\sigma_{qg}$ is the inclusive cross-section for a quark jet to split into a semi-hard quark subjet ($q$) and a semi-hard gluon subjet ($g$). We have fixed $Q=E$, the inital jet energy, and introduced the gluon energy fraction $z = E_{g}/E$. Here $\mu_{\mathrm{s}}$ is the low scale of radiation over which $\sigma_{qg}$ is inclusive. 

We write $\td \sigma_{qg}$ in the factorised form
\begin{align}
    \frac{\td \sigma_{qg}}{\td \theta \td z } = \left(1 + F_{\rm med}(z,\theta) \right)\,\frac{\td \sigma^{\mathrm{vac}}_{qg}}{\td \theta \td z }\,, \label{eq:factorisation}
\end{align}
where  $F_{\rm med}$ is the medium-induced modification, and $\td \sigma^{\rm vac}_{qg}$ is the vacuum splitting cross-section
\begin{align}
    \frac{\td \sigma^{\rm vac }_{qg}}{\td \theta \td z } = \frac{\As C_{\rm F} \sigma }{\pi} \frac{1+(1-z)^2}{z ~ \theta} + \mathcal{O}(\As^2,\theta^{0})\,,
\end{align}
with $\sigma$  the cross-section to produce the quark jet.

We are interested in illustrating how the two-point correlator can be used to identify features of the medium modification to the jet substructure. For that purpose, we use the calculation in \cite{Dominguez:2019ges,Isaksen:2020npj} (specifically eqs.~(3.6), (3.18), and (3.19) of~\cite{Isaksen:2020npj}) where the different emerging scales from the medium are properly identified using a simplified model with a static brick of length $L$ and jet-quenching parameter $\hat q$. The multiple interactions with the medium are resummed in the BDMPS-Z formalism with the harmonic oscillator approximation at leading-colour for a \emph{semi-hard splitting} where all partons propagate eikonally while undergoing colour rotations. This calculation is particularly well suited to our analysis since the angular scales can be directly read off the expression for $F_{\mathrm{med}}$ as explained in \cite{Dominguez:2019ges}. Splittings with a formation time $t_{\rm f} \sim (z(1-z)E\theta^2)^{-1}$ larger than the length of the medium $L$ are not typically expected to have a significant medium modification. For this particular model this emerges naturally as $F_{\rm med}\to 0$ exponentially for $\theta \lesssim \theta_{\rm L} \sim (EL)^{-1/2}$, in agreement with our general expectation of no modification at small angles. Generally, medium-induced radiation is also expected to dominate over vacuum when the decoherence time $t_{\rm d} \sim (\hat q \theta^2)^{-1/3}$ is smaller than the formation time $t_{\rm f}$, provided both times are smaller than the length of the medium (see also \cite{Caucal:2018dla,Caucal:2021cfb}), which can only happen above the \emph{critical angle} $\theta_{\rm c} \sim (\hat q L^3)^{-1/2}$. This too naturally emerges in our model \cite{Dominguez:2019ges}.

We can use the factorisation in Eq.~\eqref{eq:factorisation} in conjunction with the small angle $F_{\rm med}\to 0$ limit of our model to extend our calculation over the full perturbative range of $\td \Sigma^{(n)}$ and resum the small angle vacuum radiation,
\begin{align}
    &\frac{\td \Sigma^{(n)}}{\td \theta} = \frac{1}{\sigma_{qg}}\int \td z  \left(g^{(n)}(\theta,\As) + F_{\rm med}(z,\theta) \right) \frac{\td \sigma^{\rm vac}_{qg}}{\td \theta \td z }  \label{eq:masterequation} \\
    & \times z^n (1-z)^n \left(1  + \mathcal{O}\left(\As \ln \theta_{\TT{L}}^{-1}, ~\frac{\mu_{\rm s}}{z E} \right) \right) + \mathcal{O}\left(\frac{\mu_{\rm s} }{E} \right) \nonumber, 
\end{align}
where $g^{(1)} = \theta^{\gamma(3)} + \mathcal{O}(\theta)$ at fixed coupling given $\td\sigma^{\rm vac}_{qg}$ at $\mathcal{O}(\alpha_s)$. Although the expression for $g^{(n)}$ with $n>1$ is more complicated, crucially one still gets that $g^{(n)} \rightarrow 1$ as $\As \ln \theta^{-1} \rightarrow 0$.  The two new error terms are respectively for the interplay between the medium and secondary vacuum radiation, and for the semi-hard approximation used in the resummation of the medium interactions. This observable being inclusive, further energy loss experienced by the daughter partons is subleading and not enhanced by large logarithms. For a detailed discussion see \cite{Andres:2023xwr}.

There are two competing angular scales, $\theta_{\rm L}$ and $\theta_{\rm c}$, which become relevant for our analysis. We therefore expect qualitatively different behaviours depending on which one is larger. Splittings with $\theta\gtrsim\theta_{\rm c}$ are resolved by the medium, meaning that the colour coherence of the daughter partons is broken and therefore can interact with the medium independently, while splittings with $\theta<\theta_{\rm c}$ are not resolved and therefore their medium modification is much smaller. When $\theta_{\rm L}\gg\theta_{\rm c}$ all in-medium emissions are resolved by the medium and $\theta_{\rm c}$ is no longer relevant. We refer to this regime as the decoherent limit (DC). In the opposite case, $\theta_{\rm L}\ll\theta_{\rm c}$, there is an angular region in which emissions do occur inside the medium but are not fully resolved, which translates into a small but non-zero medium induced contribution. We refer to this limit as partially coherent (PC).


\begin{figure}
\includegraphics[width=0.50\textwidth]{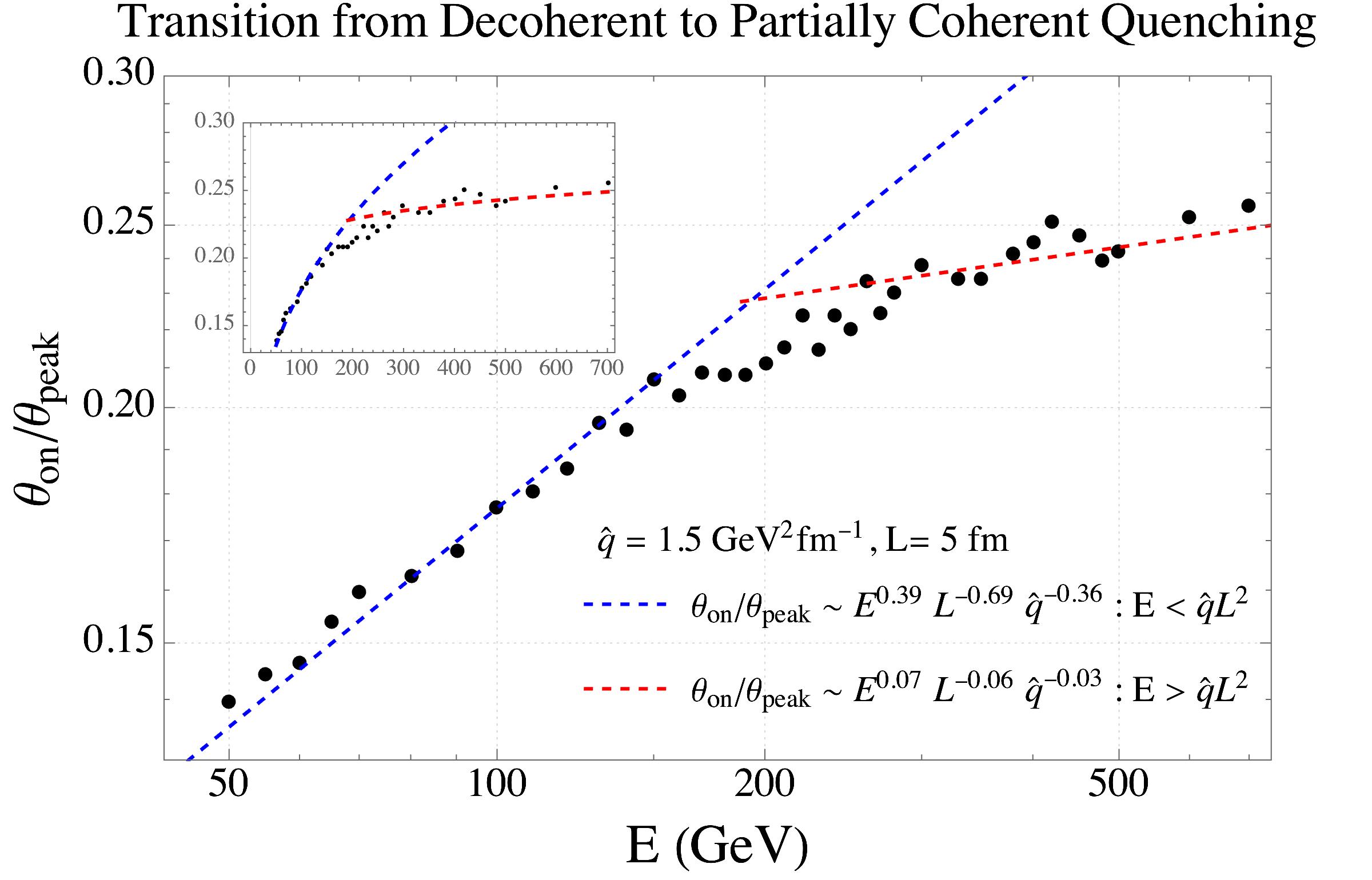}
\caption{Ratio of the peak $\theta_{\TT{peak}}$ and onset $\theta_{\TT{on}}$ angles of the EEC as a function of the jet energy.  A clear change in the power-law scaling of the ratio is visible around the critical energy, which for the shown values of $\hat{q} = 1.5$~GeV$^2$fm$^{-1}$ and $L = 5$~fm is $E_c \sim 187.5$~GeV. The embedded figure shows the same plot without the Log-Log axis. The power-law fits are provided in the legend and were extracted from a much larger data set of 242 different sets of the parameters $(E,L,\hat{q})$ from which this figure is one slice.}
  \label{fig:coherence}
\end{figure}

\emph{Numerical Evaluation.}---In Fig.~\ref{fig:EEC100GeV} we present the numerical evaluation of Eq.~\eqref{eq:masterequation} with $n=1$ and fixing $g^{(n)}=1$, which is a good first approximation away from the $\theta\rightarrow 0$ divergence. The results with $n=2$ are qualitatively the same as for $n=1$. The parameters ($E,L,\hat{q}$) have been chosen such as the left panel of Fig.~\ref{fig:EEC100GeV} corresponds to the DC while the right panel to the PC limit.  All curves satisfy our expectation of being near identical to the vacuum result for small angles and having an excess due to medium-induced radiation at large angles. The onset angle $\theta_{\rm on}$ above which the medium-jet curves deviate from the vacuum curve seems to be independent of $\hat q$  both in the DC and PC scenarios, but for larger angles the shape of the curves seems to vary in a different way for the two regimes when $\hat q$ varies. In order to quantify these observations, we performed numerical analyses for 242 other sets of the parameters with $E \in [50,700]$~GeV, $L \in [0.2,10]$~fm, and $\hat{q} \in [1,3]$~GeV$^2$fm$^{-1}$ to  extract the scaling behaviour of $\theta_{\rm on}$ and $\theta_{\rm peak}$, the position of the peak in the medium enhancement, in the PC and DC regimes. In both limits, and for both $n=1\&2$, the onset angle was found to scale as $\theta_{\rm on} \sim \theta_{\TT{L}}^{1 \pm 0.1}$, this was very robust against the extraction procedure.  In the DC limit for both $n=1\&2$, the peak position was found to scale as 
\be
\theta^{\rm DC }_{\rm peak}\sim E^{-0.86\pm 0.1} L^{0.21 \pm 0.1} \hat{q}^{0.36 \pm 0.1}\,.
\ee
In the PC limit, for both $n=1\&2$, 
\be
\theta^{\rm PC}_{\rm peak}\sim E^{-0.54\pm 0.1} L^{-0.31 \pm 0.1} \hat{q}^{0.09 \pm 0.1}\,.
\ee
Note the scalings of $\theta_{\rm on},\theta^{\rm DC}_{\rm peak },\theta^{\rm PC}_{\rm peak }$ are all dimensionally correct. This was not imposed in our fits and its emergence is an indication of their robustness. We note that the best extraction of $\theta^{\mathrm{DC}}_{\mathrm{peak}},\theta^{\mathrm{PC}}_{\mathrm{peak}}$ was achieved from $\Sigma^{(2)}$ where a slightly sharper peak is observed. We expect the change of scaling between the DC and PC regimes could be accessed experimentally with jets at different centralities at RHIC and at the LHC.

The deviation in the scaling of $\theta^{\rm DC}_{\rm peak}$ and $\theta^{\rm PC }_{\rm peak}$ is a clear indication of the emergence of decoherence/coherence in the medium-induced radiation. This can be easily seen by eye when one plots the ratio $\theta_{\rm on}/\theta_{\rm peak}$ as a function of the jet energy for fixed medium parameters $\hat q$ and $L$, largely removing the $\theta_{\TT{L}}$ dependence from $\theta^{\rm PC}_{\rm peak}$,  see Fig.~\ref{fig:coherence}. The change of regime is clearly visible in this figure at the critical energy $E_c\sim \hat q L^2$ which coincides with the condition $\theta_{\rm c}\sim\theta_{\rm L}$, thus undoubtedly signaling the emergence of a new relevant angular scale. This transition would not be possible to observe from the Lund planes only, as seen in \cite{Dominguez:2019ges,Isaksen:2020npj}, where defining specific features as the onset and peak angles would be considerably more challenging.

\begin{figure}
\includegraphics[width=0.47\textwidth]{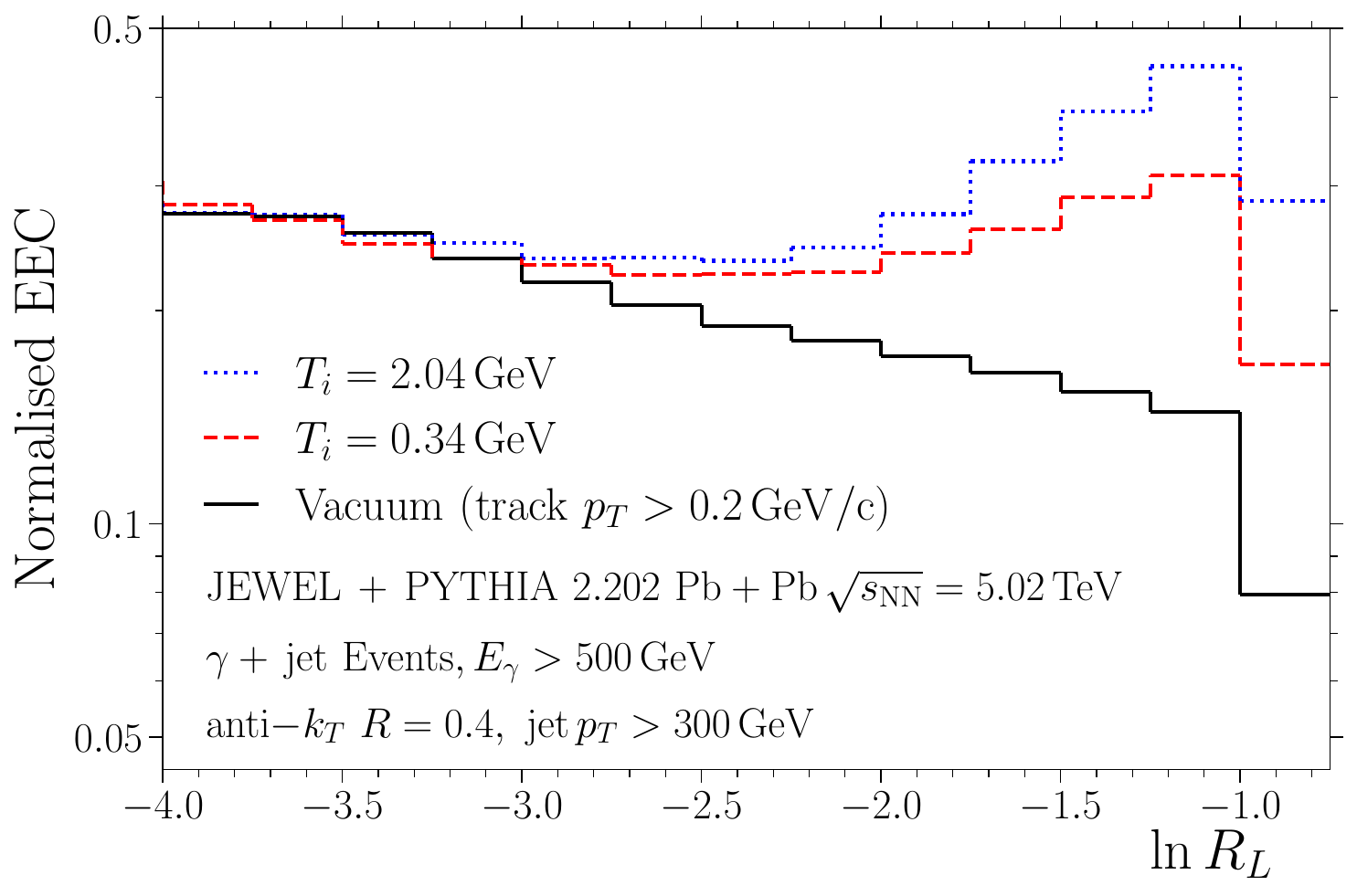}
  \caption{The EEC in \textsc{JEWEL}  with recoils for $\gamma$+jet Pb$+$Pb  events at $\sqrt{s_{\rm NN}} = 5.02$\,TeV with respect to $\ln R_L$ with $R_L=\sqrt{\Delta \phi^2+ \Delta \eta^2}$. The solid black line is the p$+$p reference, while the blue dotted and red dashed curves correspond to medium temperatures $T_i=340$\,MeV and $T_i=2.04$\,GeV, respectively.}
  \label{fig:jewel}
\end{figure}

\emph{Analysis with JEWEL.}---Having illustrated the features of the energy correlators in an idealised theoretical calculation, we now use the Monte Carlo parton shower JEWEL with recoils \cite{Zapp:2008gi,Zapp:2012ak,Zapp:2013vla,KunnawalkamElayavalli:2017hxo} to show their potential in simulations of jet-medium interactions. The EEC computed using \textsc{JEWEL}, with anti-k$_{T}$ $R=0.4$ jets recoiling off photons \cite{KunnawalkamElayavalli:2016ttl} in $\sqrt{s_{\rm NN}} = 5.02$\,TeV Pb$+$Pb collisions at $T=0.34$ GeV and $T=2.04$ GeV is shown in Fig.~\ref{fig:jewel}, and compared with the vacuum EEC. The case of $T=2.04$ GeV is unrealistically high, but is meant to illustrate the dependence on temperature. An enhancement at wide angles similar to that found in our (semi-)analytical analysis is clearly seen. We have also checked that the EEC remains robust to a 2 GeV cut on the track $p_T$, as typically used experimentally to suppress backgrounds. This illustrates the resilience of the EECs to the techniques typically employed in experimental analyses to reduce the impact of large heavy-ion backgrounds. We anticipate that the measurement of the EEC can be performed using RHIC and LHC data.

\emph{Conclusions.}---In this \emph{Letter} we have presented a completely novel approach to jet substructure in heavy-ion collisions by means of correlation functions of energy flow operators. We have analysed the specific case of the two-point correlator of a quark-gluon in-medium splitting within the BDMPS-Z framework with a harmonic oscillator potential, and showed that it is sensitive to the onset of colour coherence in the splitting process. We have found that these results can be also reproduced with other jet quenching formalisms, such as GLV \cite{Gyulassy:2000fs,Gyulassy:2000er,Wiedemann:2000za}, and other parton-medium interaction models, such as the Yukawa or Gyulassy-Wang potential \cite{Gyulassy:1993hr}. This will be shown in an upcoming publication \cite{Andres:2023xwr}. Finally, we have further validated the experimental feasibility of our analysis with the JEWEL event generator. 

Our calculations highlight in a spectacular fashion the ability of the correlators to identify the presence of dynamics at a given scale. There are two reasons that we find this particularly promising for applications to heavy-ion collisions. First, the scales of the QGP imprint themselves as large changes in the slope of the correlator, generating cusps and peaks, at scales set by dimensional analysis. Therefore, even in the absence of precise theoretical control, one can robustly identify the scales. Indeed, they are visible by eye. Second, the medium modification observed in the energy correlators \emph{cannot} be explained by modifying the quark/gluon fraction, or by biases in the jet $p_T$ selection. These effects only modify the correlator distribution \emph{logarithmically} (through modifications of the anomalous dimensions), without introducing changes of sign in the angular dimension, preventing the generation of cusps and peaks. To our knowledge, this is a unique feature of the energy correlators, making them an ideal observable to probe the complicated multi-scale dynamics of the QGP. For other recent approaches aiming at resolving the time structure of the QGP, see~\cite{Apolinario:2017sob,Andres:2019eus,Apolinario:2020uvt}.

While we demonstrated the potential of our approach for the specific case of the EEC for in-medium massless jets, there are many generalizations of our philosophy which can shed further light on the dynamics of the QGP. A natural generalisation is to higher point correlators, which have been shown to provide interesting insights into vacuum jets \cite{Chen:2019bpb,Chen:2020adz,Holguin:2022epo,Komiske:2022enw,Chen:2022swd}. Correlators involving heavy quarks are also particularly interesting in the medium, since they introduce another intrinsic scale, and are not often pair produced, allowing them to be tracked through the medium (for recent interesting uses of heavy quarks, see e.g.~\cite{Cunqueiro:2018jbh,ALICE:2021aqk,Apolinario:2022vzg,Ke:2022gkq}). Finally, we have focused on the perturbative region of the EEC, but it would be also interesting to study medium modifications to the hadronisation transition, which has already been analysed in vacuum \cite{Komiske:2022enw}.

We believe that understanding \emph{how} the dynamics of the QGP can be robustly extracted from jet substructure observables is the first step towards achieving new levels of theoretical control over jet substructure observables in heavy-ion collisions. In this \emph{Letter} we have shown how this can be achieved using energy correlators, with the added benefit of opening the doors to a wealth of theoretical techniques recently developed to study these observables. We look forward to a rich program unravelling the structure of the QGP with energy correlators.

\emph{Acknowledgements.}--- We thank Leticia Cunqueiro, Cristian Baldenegro Barrera, Carlos A. Salgado, Liliana Apolinario, Jasmine Brewer, Andrew Larkoski, Xin-Nian Wang, Laura Havener, Alba Soto-Ontoso, Peter Jacobs, Emery Sokatchev, Alexander Zhiboedov, Kyle Lee, Helen Caines, Ivan Vitev for helpful/encouraging discussions. We thank Gregory Soyez for a useful comment made during the BOOST workshop. This work is supported in part by the GLUODYNAMICS project funded by the ``P2IO LabEx (ANR-10-LABX-0038)'' in the framework ``Investissements d’Avenir'' (ANR-11-IDEX-0003-01) managed by the Agence Nationale de la Recherche (ANR), France. This work is also supported by European Research Council project ERC-2018-ADG-835105 YoctoLHC; by Maria de Maetzu excellence program under project MDM-2016-0692 and CEX2020-001035-M; by Spanish Research State Agency under project PID2020-119632GB-I00; and by Xunta de Galicia (Centro singular de investigación de Galicia accreditation 2019-2022), by European Union ERDF, and by OE - Portugal, Funda\c{c}\~ao para a Ci\^encia e Tecnologia (FCT) under project EXPL/FIS-PAR/0905/2021. C.A. has received funding from the European Union’s Horizon 2020 research and innovation program under the Marie Sklodowska-Curie grant agreement No 893021 (JQ4LHC). I.M. is supported by start-up funds from Yale University. RKE was supported in part by the US DOE under award number DE-SC004168.

\bibliography{refs.bib}{}
\bibliographystyle{apsrev4-1}
\newpage
\onecolumngrid
\newpage

\end{document}